\documentclass[aps,pre,twocolumn,groupedaddress,floatfix]{revtex4}
\usepackage{graphicx} 
\usepackage{graphics}
\usepackage{epsfig}
\usepackage{amsmath}
\usepackage[svgnames]{xcolor}
\usepackage{todonotes}

\usepackage{outlines} 

\begin{document}
\title{The Excluded Area of Superellipse Sector Particles}

\author{Kellianne Kornick}

\altaffiliation [Rochester Institute of Technology, School of Physics and Astronomy]{}

\author{Ted Brzinski}

\altaffiliation [Haverford College, Department of Physics and Astronomy]{}

\author{Scott V. Franklin}

\altaffiliation [Rochester Institute of Technology, School of Physics and Astronomy]{}


\begin{abstract} \noindent 
Superellipse sector particles (SeSPs) are segments of superelliptical curves that form a tunable set of hard-particle shapes for granular and colloidal systems. SeSPs allow for continuous parameterization of corner sharpness, aspect ratio, and particle curvature; rods, circles, rectangles, and staples are examples of shapes SeSPs can model. We investigate the space of allowable (non-overlapping) configurations of two SeSPs, which depends on both the center-of-mass separation and relative orientation. Radial correlation plots of the allowed configurations reveal circular regions centered at each of the particle's two endpoints that indicate configurations of mutually-entangled particle interactions. Simultaneous entanglement with both endpoints is geometrically impossible; the overlap of these two regions therefore represents an excluded area in which no particles can be placed regardless of orientation. The regions' distinct boundaries indicates a translational frustration with implications for the dynamics of particle rearrangements (e.g. under shear). Representing translational and rotational degrees of freedom as a hypervolume, we find a topological change that suggests geometric frustration arises a phase transition in this space.  The excluded area is a straightforward integration over excluded states; for arbitrary relative orientation this decreases sigmoidally with increasing opening aperture, with sharper SeSP corners resulting in a sharper decrease. Together, this work offers a path towards a unified theory for particle shape-control of bulk material properties.
\end{abstract}

\maketitle

\section*{Introduction}

The importance of particle shape has long been recognized in the studies of liquid crystals,\cite{Onsager1949, FR950} colloids \cite{Glotzer2007,PhysRevE.94.022124,Sacanna2011} and granular materials. \cite{PhysRevLett.108.208001,PhysRevLett.112.148301,C5SM02335A} With more complicated, non-spherical, shapes comes a more complicated coupling between relative particle position and orientation. The excluded area encodes this relationship between position and orientation, and is critical to the formulation of theories for phase transitions and critical packing fractions. Boubl{\'\i}k \cite{1974MolPh..27.1415B} investigated the thermodynamics of convex molecular fluids; Onsager\cite{Onsager1949} analytically calculated the excluded volume of spherocylinders to explain phase transitions in liquid crystals. Philipse \cite{Philipse1996} and Desmond \cite{Desmond2006} used similar calculations, analytic and numeric, to explain packings of long, thin granular materials. More recently, study has turned to a variety of non-convex shapes, including U-shaped staples \cite{PhysRevLett.108.208001}, bent-core liquid crystals~\cite{goodby2014handbook}, and annular-sector colloidal particles~\cite{PhysRevE.94.022124,doi:10.1021/jacs.5b10549}.  Geigenfeind and de las Heras \cite{Geigenfeind2019PrincipalCA} used principal component analysis to study excluded area as a function of relative particle orientation, with antiparallel orientations minimizing the excluded area and particle elongation the most significant shape-factor. 

\begin{figure}[b!]
    \centering
    \includegraphics[width=\linewidth]{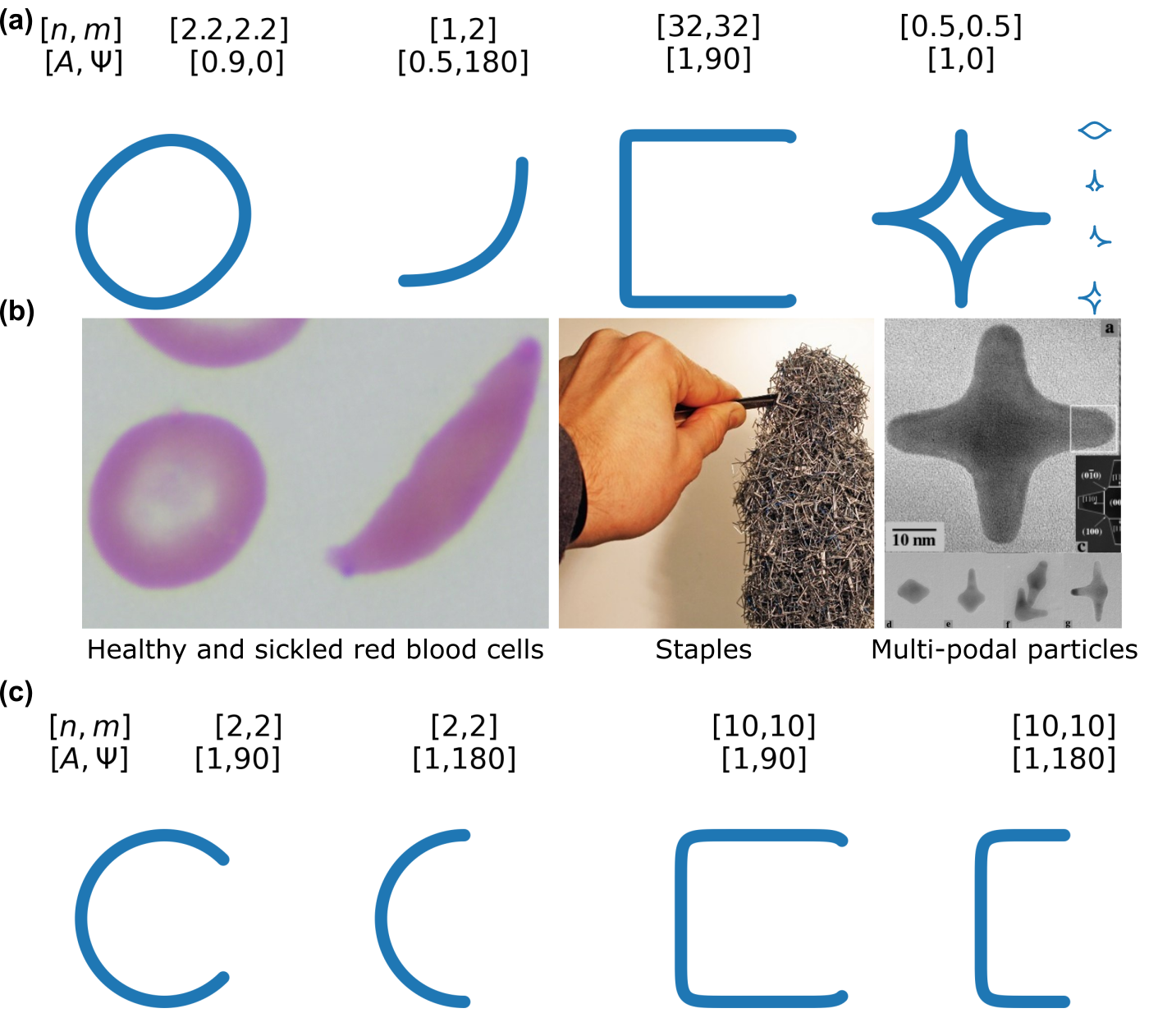}
    \caption{(a) Examples of SeSP parameterization producing shapes reminiscent of (b) (left-to-right) red blood cells of healthy and atypical shape~\cite{bcells}, staples~\cite{PhysRevLett.108.208001}, and multi-podal nano-crystaline particles~\cite{mpparts}. (c) Examples of SeSPs with $n=m\geq2$, used in the present study.}
    \label{fig:Examples}
\end{figure}

In granular materials, the critical volume fraction at which random packings form depends strongly on the symmetries of the constituent particles~\cite{Donev990}.
Particle anisometry also results in an increased angle of repose~\cite{BEAKAWIALHASHEMI2018397, PhysRevE.82.011308}, even exceeding 90$^\circ$ for particles with extreme aspect ratio~\cite{PhysRevE.82.011308}, indicating a bulk tensile strength that resists extension even though the particles themselves are purely repulsive~\cite{doi:10.1002/9781119220510.ch17}. This {\it geometric cohesion} results from particle entanglement that constrains orientational rearrangements. 
Subsequent studies found geometric entanglement in staple-shaped particles, which entangle in a manner consistent with a random contact model~\cite{PhysRevLett.108.208001}. Packings of granular staples, star-shaped, both rigid and flexible rods, and z-shaped particles have also exhibited tensile strength~\cite{PhysRevE.101.062903}, with relevance for aleatory design~\cite{doi:10.1063/1.5132809} and strengthening granular materials under strain~\cite{JaegerShear,Bares2019,PhysRevLett.120.088001}. Banana-shaped or bent-core rods are of broad interest to the liquid crystal community because of their rich liquid crystalline phase-space~\cite{Yangeaas8829,PhysRevX.4.011024,FR950,RevModPhys.90.045004}. Semi-circular particles are a compelling, quasi-2D model system for the study of homodimerization and chirality-driven phase separation~\cite{PhysRevE.94.022124,doi:10.1021/jacs.5b10549}, and entropy approaches have been applied successfully to colloidal crystals of a variety of shapes~\cite{Gengeaaw0514, Glotzer2007}. Recently, \cite{PhysRevE.102.042903} used Monte-Carlo techniques to investigate packings of hard, circular arcs, analytically identifying densest configurations which can then be compared with simulations to identify the likelihood of their appearing in bulk packings. While many of these prior observations feature particles with qualitatively similar geometric features, there is no unifying framework to describe particle geometry as a general parameter space.

We present here a study of a generalized class of particles --- Superellipse Sector Particles (SeSPs) --- based on a parameterized formulation of superellipses, or Lam\'{e} curves, a family of closed, two-dimensional curves which allow for continuous variation of angularity, convexity and aspect ratio. By varying the appropriate parameters, SeSPs can approximate many of the geometries studied in prior studies (illustrated in Fig.~\ref{fig:Examples}, including disks, ellipses, rods, U-shaped, Z-shaped and other convex, non-convex, or chiral particles. Thus, SeSPs are a compelling model system for the study of generalized particle shape-effects in particulate systems.  In this work, we study the allowable configurations in two dimensions of  two identical SeSPs with random orientation and center-of-mass separation. We observe pronounced, sharply delineated features in the distribution of allowed states that corresponds to mutually entangled states. We find, tantalizingly, a topological change in the hypervolume of excluded, pairwise configurations, hinting at the possible existence of a phase transition due to the onset of an associated geometric frustration. 

\section*{Methods}

Superellipsoids (also known as Lamé curves) are defined as the set of points $\left| \frac{x}{a} \right|^m + \left| \frac{y}{b} \right|^n = 1$ , which we parameterize as
\begin{equation}
\begin{bmatrix}
    x(\theta) \\
    y(\theta) \\
         \end{bmatrix} = 
\begin{bmatrix}
    \left|\cos\theta\right|^{2/n} \; A \; \textrm{sign}(\cos\theta) \\
    \left|\sin\theta\right|^{2/m} \; \textrm{sign}(\sin\theta) \\
        \end{bmatrix}.
\label{Eqn:para}
\end{equation}
$a$ and $b$ function as semi-major and semi-minor axes, with $A\equiv a/b$ defining the particle aspect ratio, while $m$ and $n$ are the superellipse degrees which, together, tune the particle curvature.
Values of $m$ or $n$ below 2 produce non-convex shapes, with sharp corners resulting as $m,n\rightarrow \infty$ and for all $m,n\leq1$.
$\theta$ is the parametric variable, which uniquely reduces to the polar angle for the case of $m=n=2$.
Segments are obtained by selecting a continuous subset $\hat{\theta} = [\theta_{min}, \theta_{max}]$ within this range. While $\Delta\theta$ is not the polar angle, $\Delta\theta=\theta_{max}-\theta_{min}=2\pi$ always produces a closed curve, and  $\Delta\theta=n\pi/2$ always produces a segment which subtends a solid angle of $\pi$.

The infinitesimally thin SeSPs in this work are thus defined by five parameters: the aspect ratio $A$, degrees $m$ and $n$, and segment ends $\theta_{min}$ and $\theta_{max}$. $A=1, m=n=2$ forms a special case known as an {\it annular sector particle} (ASP) that has been studied previously in colloidal systems~\cite{Wang_Mason_JACS}. We also define the \emph{opening} $\Psi=2\pi-\left(\theta_{max}-\theta_{min}\right)$. For this work we  center the opening aperture around $\theta=0$, with $\theta_{min} \equiv -\theta_{max}$ and define the SeSP orientation with a unit vector $\hat \beta$  pointing from particle's center of curvature through the midpoint of the opening aperture ($\theta=0$ direction). We numerically construct our particles and check for intersections using the procedure described in Ref.~\cite{Pilu1995}.

The excluded area of a particle is defined as the fraction of area forbidden to a particle due to the existence of another. A straightforward method for determining this (e.g. \cite{PhysRevE.67.051302}) is a Monte-Carlo simulation which proceeds as follows:
(1) a particle is fixed at the origin and (2) a second particle is placed at a random location and orientation and checked for overlap with the first. The second particle is then moved to a new (random) location and orientation and the process repeats. The excluded area is the fraction of intersecting placements multiplied by the area of the possible space
\begin{equation}\label{eq:Ae}
    A_{ex} = \left( \frac{N_{overlap}}{N_{total}}\right)  A_{box},
\end{equation}
where $N_{overlap}$ is the number of states in which particles overlap with the particle at the origin, $N_{total}$ is the total number of simulated particles, and $A_{box}$ is the size of the simulation box.  In addition to calculating the excluded area, we also record the location and relative orientation of the second particle. $10^5$  particle placements are made for each parameter set.

\section*{Results}

\subsection*{Mapping Allowed Configurations}

\begin{figure}[!h!t!b]
  \centering
  \includegraphics[width=\linewidth]{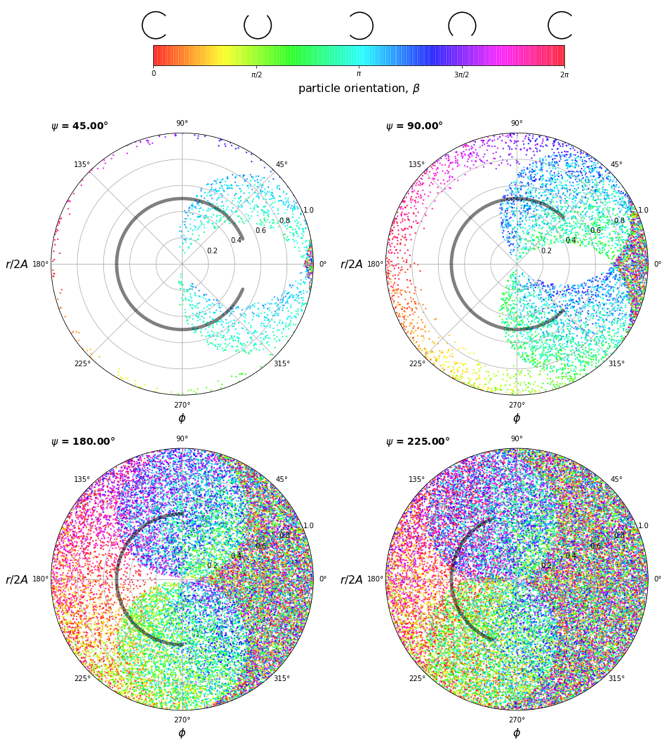}
  \caption{Allowable configurations between two ASP particles for four opening angles, where the fixed ASP is shown on each plot, and dots show the permitted states for the center of curvature of the second particle. Color representation relative particle orientation, with bluer states being anti-parallel, and redder states parallel.}
  \label{fig:ASP_radial}
\end{figure}

The excluded area defined by Eq.~\ref{eq:Ae} is an incomplete description of the spatial extent of the influence of a SeSP because exclusion depends both on relative orientation and relative position. The accessibility of relative positions has a non-trivial dependence on relative orientation that the traditional orientational correlation function fails to capture. Geigenfeind and de las Heras investigated the excluded area as a function of fixed orientation; to explore the complexity that arises when all orientations are possible we plot the allowed configurations of placed particles relative to a fixed particle (placed at the origin) in Fig.~\ref{fig:ASP_radial}. Dots mark allowed placements of a second particle, and regions with no dots are excluded regions. Dots are placed at the centers of curvature of permitted states, and the color represents the allowed orientation relative to the fixed reference SeSP.

In general allowed placements fall into four regions of interest, delineated by sharp boundaries. 
First, an \emph{unexcluded} region in which a particle can be placed without any restriction on orientation. For the case of $n=2$, this corresponds to particle separation of $\nabla r\geq d$ where $d$ is the diameter of the SeSP plus a cusped region in front of the reference SeSP's opening. Second, a \emph{spooning} region behind the arc of the reference SeSP where particles can only assume orientations such that their opening angles pointing towards the center of the fixed particle. Third, two circular \emph{entanglement} regions, centered on the reference SeSP's endpoints, with diameter equal to that of a circle that intersects with each endpoint and the midpoint of the reference SeSP. In these regions, only mutually entangled states are allowed, such that the SeSPs interpenetrate, i.e. each particle intersects with the convex hull of the other.
Finally, an \emph{excluded} region close to the reference SeSP in which no particles may be placed regardless of orientation. The boundary of the excluded region is the envelope of the family of curves defined as circles having a radius of $r = d$ with their centers lying on the arc of the SeSP. 
For $\Psi\geq\pi$, the excluded region is contiguous, but for $\Psi\leq\pi$, there is an additional excluded region where the entanglement regions overlap.

\begin{figure}[!h!t!b]
  \centering
  \includegraphics[width=\linewidth]{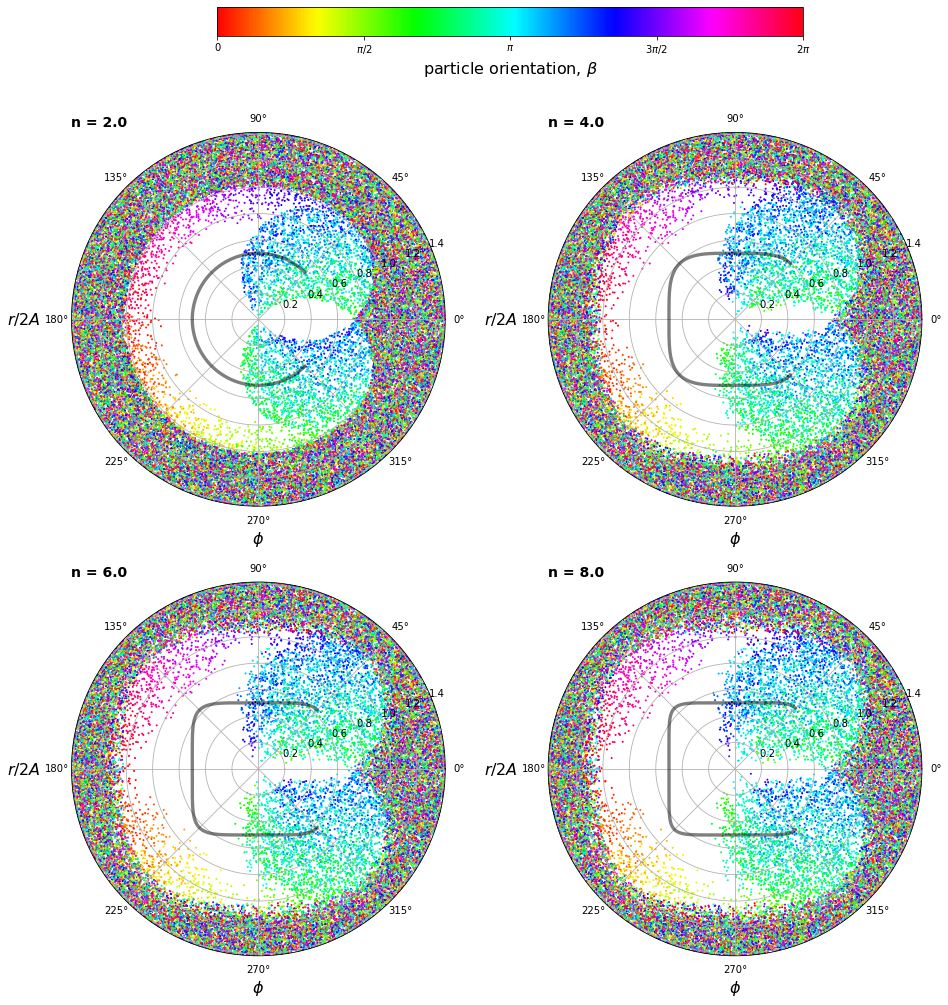}
  \caption{Allowable configurations between two SeSP particles for four different SeSP exponents. All SeSPs have an opening angle of 90 degrees. The fixed SeSP is shown at the center of each plot, and dots are defined in the same way as in Fig. \ref{fig:ASP_radial}. Regions of correlation between relative positions and orientation become smeared as the corner of the SeSP becomes sharper.}
  \label{fig:SeSP_radial}
\end{figure}

Next we generalize to SeSPs with larger exponents $m,n$. 
Fig.~\ref{fig:SeSP_radial} shows allowed configurations for SeSPs with four different exponents, all with opening angles of $\Psi = \pi/2$. While we still observe the distinct and sharply delineated regions, their shapes become distorted. For SeSPs of higher power (sharper corners, more box-like), the shape of the excluded area retains its lobed boundary on the right side of each subplot, but the left side is distorted to have sharp corners for states in which the two SeSPs may only lie edge-to-edge. For all powers $m,n>2$, we observe the the topological change in the shape of the excluded region when the entangled regions begin to overlap for $\Psi<\pi$.The spooning region is constricted along parts of the SeSP with a large radius of curvature, and expanded near regions with small radius of curvature. The entangled regions become larger, with fewer accessible orientations near their edges, leading to a fuzzy or blurred appearance in Fig.~\ref{fig:SeSP_radial}. 

\subsection*{Excluded Area}
\begin{figure}[htp]
  \centering
  \includegraphics[width=\linewidth]{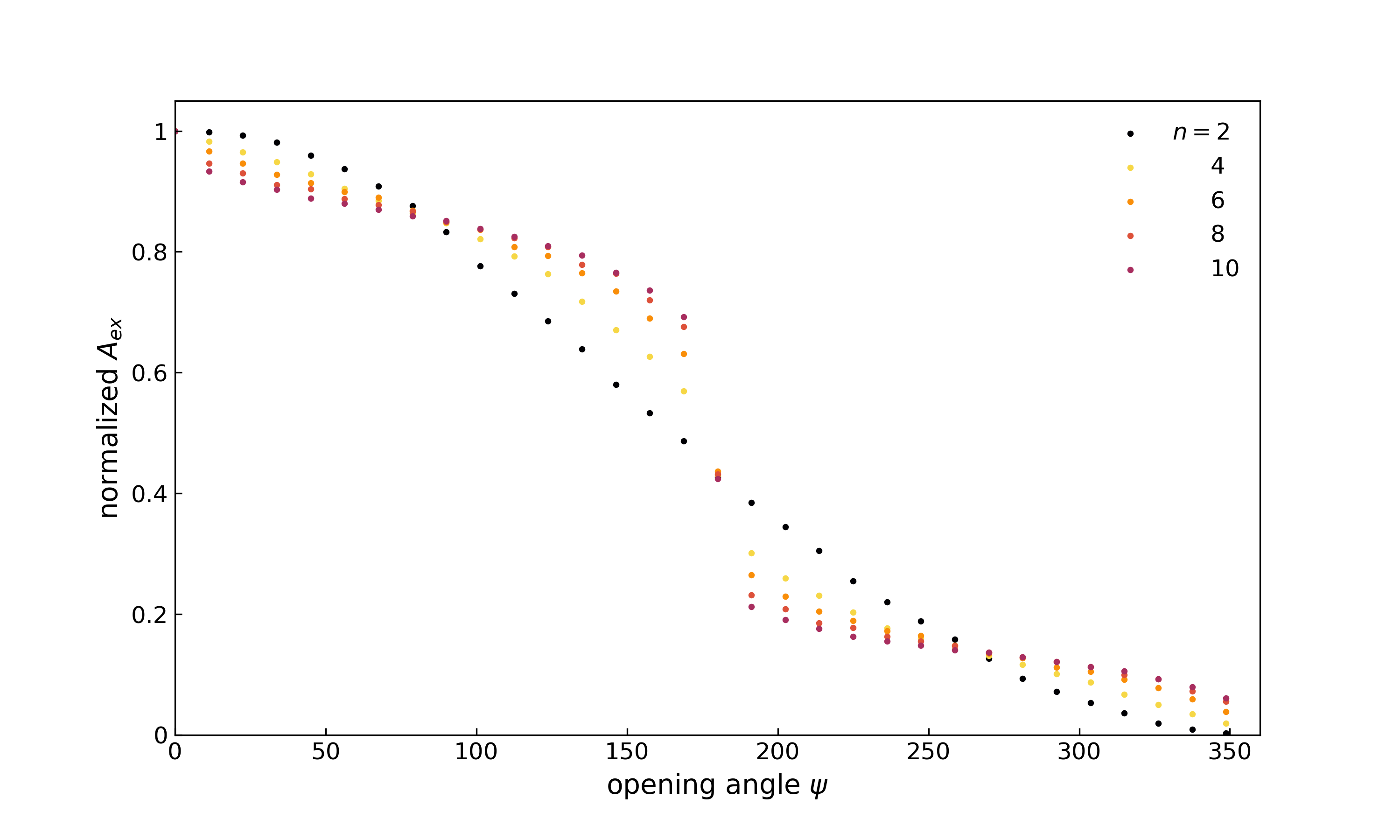}
  \caption{SeSP particles exclude less area from each other as the opening aperture increases. As $\Psi \to 2\pi$, since the ASPs are constructed from closely-spaced circles, there is at least one constituent circle remaining, accounting for the small, non-zero excluded area right before $\Psi = 2\pi$. Areas are normalized by the excluded area with zero opening angle, or the excluded area of a disk.}
  \label{fig:A_ex_tot_ASP}
\end{figure}

Finally, we calculate the excluded area using the Monte-carlo method described above. Figure \ref{fig:A_ex_tot_ASP} shows the total excluded area of a SeSP as a function of opening aperture $\Psi$ for the special case of $m=n=2$. As expected, the excluded area decays from 1 (normalized by the excluded area $\pi D^2$ of a full circle) sigmoidally to zero as the opening aperture is increased to $2\pi$. This behavior is characteristic of all SeSPs of $n>2$, with the sigmoid sharpening to a sloped step as $n\rightarrow\infty$.

\begin{figure*}
    \centering
    \includegraphics[width=\linewidth]{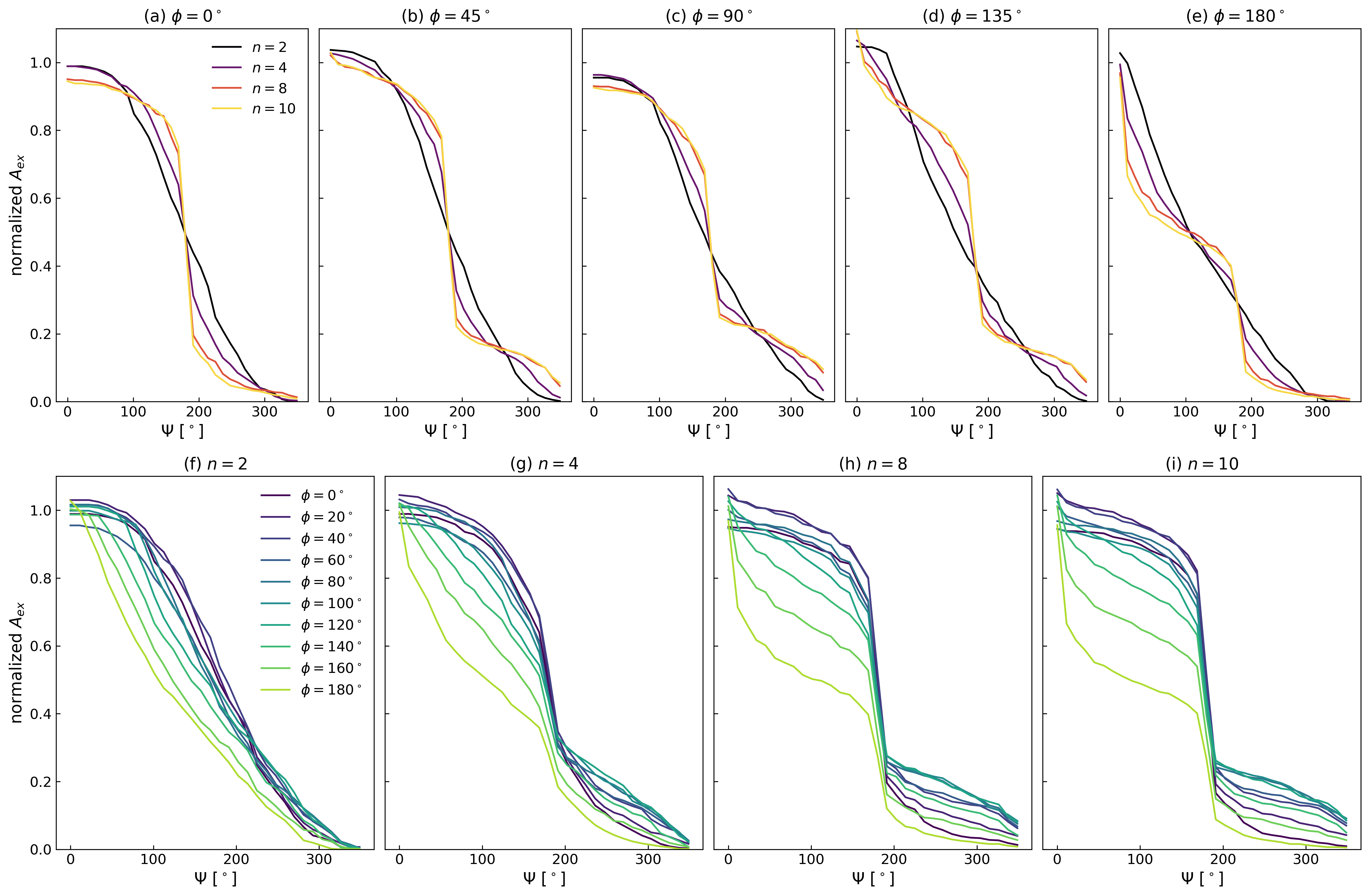}
    \caption{Normalized excluded area plotted as a function of opening angle for (a)-(e) fixed relative orientation $\phi=0^\circ,45^\circ,90^\circ,135^\circ,180^\circ$ respectively with power $n$ indicated by line color, and (f)-(i) fixed power $n=2,4,8,10$ respectively, with relative orientation indicated by line color.}
    \label{fig:Rev2PITA}
\end{figure*}

Figure \ref{fig:Rev2PITA} shows how the total excluded area depends on relative orientation $\phi$ and SeSP exponent $n$. Several aspects are noteworthy. First, the excluded area decreases monotonically with orientation, reaching a minimum at $180^\circ$ for all particle shapes, as seen in Fig.~\ref{fig:Rev2PITA} (f-i) and consistent with \cite{Geigenfeind2019PrincipalCA}. Second, there is a dramatic change in the shape of $A_{ex}$ vs $\Psi$ at $\phi=90^\circ$. Below $\phi=90^\circ$, there is a low-$\Psi$ plateau followed by a monotonic decay; above $\phi=90^\circ$, there is an immediate sharp decay at low-$\Psi$. This change from plateau to decay corresponds to the ability of one particle to penetrate through the opening aperture of the other, or equivalently, the decay of the excluded region at the intersection of the entangled regions as shown in Fig~\ref{fig:ASP_radial}. The qualitative change from plateau to decay is most pronounced for $n>>2$, where increasingly (with $n$) flat particle sides give rise to an intermediate plateau. For large values of $\Psi$ (>$270$) the system behavior approaches that of rod-like particles.

\section*{Conclusion}

We have presented a new, generalizable framework for approximating an extraordinarily broad class of 2D particle shapes. We then use this framework to study the complex interaction between orientation and center-of-mass separation of concave particles, revealing areas of mutual entanglement whose overlap defines forbidden zones in the radial distribution of allowed states.
The spatio-orientational distribution of allowed states undergoes a topological change at $\Psi=180^\circ$, below which the forbidden zone vanishes, hinting at qualitatively different particle scale structure or dynamics in many-body dispersions of SeSPs.
The nature of the topological change suggests that additional geometric frustration will exist for SeSPs of $\Psi<180^\circ$. For $\Psi>180^\circ$, direct trajectories exist between the different entanglement regions, whereas for $\Psi<180^\circ$, trajectories between entanglement regions must include states in the unexcluded region or spooning region. We thus hypothesize the existence of a new glass-like phase for materials comprised of $\Psi<180^\circ$ SeSPs as a consequence of frustration due to mutual entanglement rather than structural confinement. Future simulations and experiments are needed to explore the existence of a phase transition corresponding to this state.

We have also numerically calculated the excluded area $A_{excl}$ for SeSPs of $n=m\geq2$ and $A=1$, a quantity of fundamental interest in developing theoretical predictions for states of matter governed by particle-scale geometric constraints. This calculation recaptures the previously observed minimum in $A_{excl}$ with orientation and reveals the impact of sharp corners and flat sides that accompany large values of the exponent parameter $n$. Together, the framework and accompanying tools introduced above offer a path towards a unified theory for particle shape-control of bulk material properties in contrast to the more empirical cataloging approach that has characterized most prior work.

Future work in this area includes measuring the critical random (loose and close) packing fractions and their correlation to the calculated excluded area, how the distribution of allowed, pair-wise configurations compare with the distribution of those observed in many-body packings, and how to express generalized shape characteristics like sphericity in terms of the SeSP parameters. Finally, our particle shape formulation may also be extended to shapes with different symmetries~\cite{https://doi.org/10.3732/ajb.90.3.333}, or to higher dimensions (by taking sections of superellipsoidal shells).

\begin{acknowledgments}
We acknowledge useful support from members of the RIT School of Physics and Astronomy and the Soft Matter Research Group.
\end{acknowledgments}

\bibliography{bib}

\end{document}